\documentstyle[12pt]{article}

\begin{document}
\def\thebibliography#1{\section*{REFERENCES\markboth
 {REFERENCES}{REFERENCES}}\list
 {[\arabic{enumi}]}{\settowidth\labelwidth{[#1]}\leftmargin\labelwidth
 \advance\leftmargin\labelsep
 \usecounter{enumi}}
 \def\newblock{\hskip .11em plus .33em minus -.07em}
 \sloppy
 \sfcode`\.=1000\relax}
\let\endthebibliography=\endlist

\begin{titlepage}

\hoffset = .5truecm
\voffset = -2truecm

\centering

\null
\vskip -1truecm
\rightline{\small \it To pre-print: submit the LaTeX file to the\ \ \ \ }
\rightline{\small \it publications office with completed\ \ \ \ }
\rightline{\small \it authorization form.\ \ \ \ }
\vskip 1truecm

{\normalsize \sf \bf United Nations Educational, Scientific and Cultural
Organization\\
and the\\
International Atomic Energy Agency\\}
\vskip 1truecm
{\huge \bf
INTERNATIONAL CENTRE\\
FOR\\
THEORETICAL PHYSICS\\}
\vskip 3truecm

{\LARGE \bf
Interlinked 3D and 4D 3-Quark Wave Functions In A Bethe-Salpeter Model
}\\
\vskip 1truecm

{\large \bf
A.N.Mitra
\\}

\vskip 8truecm

{\bf MIRAMARE--TRIESTE\\}
January 1991

\end{titlepage}

\hoffset = -1truecm
\voffset = -2truecm

\title{\bf
Interlinked 3D and 4D 3-Quark Wave Functions In A Bethe-Salpeter Model
}

\author
{\bf
A.N.Mitra\thanks{Short term visitor; Permanent address for 
correspondence: 244 Tagore Park, 
Delhi-110009, India; e.mail: csec@doe.ernet.in (subject:a.n.mitra)} 
}
\normalsize International Centre for Theoretical Physics, Trieste 34100, 
{\bf Italy}

\date{13th November 1996}
\newpage

\maketitle

\begin{abstract}
Using the method of Green's functions within the framework of a 
Bethe-Salpeter formalism characterized by a pairwise $qq$ interaction 
with a 3D support to its kernel (expressed in a Lorentz-covariant manner), 
the 4D BS wave function for a system of three identical relativistic spinless 
quarks is reconstructed from the corresponding 3D form which satisfies a 
{\it fully connected} 3D BSE. This result is a 3-body generalization of a 
similar interconnection between the 3D and 4D 2-body wave functions that had 
been found earlier under identical conditions of a 3D support to the 
corresponding BS kernel, using the ansatz of Covariant Instaneity for the 
pairwise $q\bar q$ interaction. The generalization from spinless to fermion 
quarks is straightforward.  

\end{abstract}

\newpage


\section{Introduction: Statement of Problem and Summary of Results }

The problem of connectedness in a three-particle amplitude has been in 
the forefront of few-body dynamics since Faddeev's classic paper [1] 
showed the proper perspective by emphasizing the role of the 2-body 
T-matrix as a powerful tool for achieving this goal. The initial stimulus 
in this regard came from the separable assumption to the two-body 
potential [2] which provided a very simple realization of such 
connectedness via the T-matrix structure envisaged in [1], a result that 
was given a firmer basis by Lovelace [3]. An alternative strategy for 
connectedness in an $n$-body amplitude was provided by Weinberg [4] through 
graphical equations which brought out the relative roles of T- and V- 
matrices in a more transparent manner. As was emphasized in both [3] and 
[4], an important signal for connectedness in the 3-body (or $n$-body) 
amplitude is the {\it absence of any delta function} in its structure, 
either explicitly or through its defining equation. This signal is valid 
irrespective of whether or not the V- or the T- matrix is employed for 
the said dynamical equation.
	The above results were found for a non-relativistic n-body 
problem within a basically 3D framework whose prototype dynamics is the 
Schroedinger equation. For the corresponding {\it relativistic} problem 
whose typical dynamics may be taken as the Bethe-Salpeter equation (BSE) 
with pairwise kernels within a 4D framework, it should in principle be 
possible to follow a similar logic, using the language of Green's 
functions with corresponding diagramatic representations [4], leading to 
equations free from delta functions. However there are other {\it physical}   
issues associated with a 4D support to the BS kernel of a `confining' type, 
such as O(4)-like spectra [5], while the data [6] exhibit only O(3)-spectra.  
To handle this issue in a realistic and physically plausible manner, 
there have been many approaches in the literature (which hardly need to 
be cited) that are centred on a basically "instantaneous" approximation to 
the (pairwise) two-particle interaction. In the spirit of this general 
philosophy, and keeping close to the {\it observational} features of the 
hadron mass spectra [6], a concrete `two-tier' strategy [7] had been 
proposed by incorporating the physical content of the Instantaneous 
Approximation, albeit treated {\it covariantly} [8], wherein the 3D 
reduction of the original 4d BSE would serve for the dynamics of the 
spectra, while the reconstructed 4D wave function would be appropriate 
for applications to various transition amplitudes by standard Feynman 
techniques for 4D quark loop integrals [7,8]]. This philosophy found [8]
a precise mathematical content through the ansatz of Covariant Instantaneity  
(CIA for short) which gives a {\it 3D support} to the BSE kernel. This 
ansatz yields a complete equivalence between the 3D and 4D forms of the 
BSE, viz., not only is the 4D form exactly reducible to the 3D form, but 
conversely the 4D BS vertex function ${\Gamma}$ is fully expressible as a 
simple product of only 3D quantities, viz., $Dx{\phi}$, where $D$ and 
${\phi}$ are both 3D denominator and wave functions respectively, 
satisfying a relativistic Schroedinger-like equation [8]. (The ansatz 
of a 3D support to the BS kernel was also advocated by Pervushin and 
collaborators [9], but under a different philosophy from the two-tier 
point of view enunciated in [8], so that the feature of 3D-4D 
interconnection was apparently not on their agenda). A comparative 
assessment of this method vis-a-vis more conventional ones employing 
the BSE has been given elsewhere [10]. 
	One may now ask: Does a similar 3D-4D interconnection exist in 
the corresponding BS amplitudes for a {\it three-body system} under 
the same conditions of 3D support to the pairwise BS kernel? This 
question is of great practical value  since the 3D reduction of the 
4D BSE (under conditions of a 3D support for the pairwise kernel) 
already provides a a {\it fully connected} integral equation [11], 
leading to an approximate analytic solution (in gaussian form) for 
the corresponding 3D wave function, as a by-product of the main 
results on the baryon mass spectra [11]. Therefore a reconstruction 
of the 4D $qqq$ wave (vertex) function in terms of the corresponding 
3D quantities will open up a vista of applications to various types 
of {\it transition amplitudes} involving $qqq$ baryons, just as in 
the ${\bar q}q$ case [8]. This is the main purpose of the present 
investigation, with three identical spinless particles for 
simplicity and definiteness, which however need not detract from 
the generality of the singularity structures. The answer is found 
to be in the affirmative, except for the recognition that a 3D 
support to the pairwise BS kernel implies a truncation of the 
Hilbert space. Such truncation, while still allowing an unambiguous 
reduction of the BSE from the 4D to the 3D level, nevertheless leaves 
an element of ambiguity in the {\it reverse direction}, viz., 
from 3D to 4D. This limitation for the reverse direction is quite 
general for any $n$-body system where $n > 2$; the only exception is 
the case of $n = 2$ where both transitions are reversible without                   
any extra assumptions (a sort of degenerate situation). The extra 
assumption needed to complete the reverse transition in its simplest 
form is facilitated by some 1D delta-functions which however have 
nothing to do with connectedness [3,4] of an $n$-body amplitude (see 
Sec.4 for a formal demonstration). 
	The paper which makes use of Green's function techniques to 
formally derive the results stated above, is organized as follows. 
In Sec.2 we first derive the 3D-4D interconnection [8] at the level of 
the {\it Green's function} for a ${\bar q}q$ system, under conditions of 
a 3D support for the BS kernel, whence we reproduce the previously 
derived result [8] for the corresponding BS wave functions in 3D and 4D 
forms. In Sec.3, starting with the BSE for the 4D Green's function for 
{\it three} identical spinless quarks ($q$), when the $qq$ subsystems 
are under pairwise BS interactions with 3D kernel support, this 4D  BS
integral equation is reduced to the 3D form by integrating w.r.t. 
{\it two} internal time-like momenta, and in so doing, introducing 3D 
Green's functions as double integrals over two time-like internal 
momenta. The resulting 3D BSE has a fully connected structure, free from 
delta-functions, as anticipated from an earlier analysis with 3D BS 
wave functions [11,12]. With this 3D BSE as the check point, Sec.4 is 
devoted to the task of reconstructing the full 4D Green's function in 
terms of its (partial) 3D counterparts, so as to satisfy exactly the 
above 3D BSE, {\it after} integration w.r.t. the relevant time-like 
momenta. (In doing these various manipulations, the inhomogeneous parts 
of the various Green's functions will not be kept track of, since we 
shall be mainly concerned with their bound state poles). Sec.5 concludes 
with a discussion of this result, including the technical issues arising 
from the inclusion of spin, together with a comparison with contemporary 
approaches to the {\it relativistic} $qqq$ problem. 

\section{3D-4D Interconnection For the ${\bar q}q$ System}

	If the BSE for a spinless ${\bar q}q$ system has a 3D support for 
its kernel $K$ in the form $K({\hat q,\hat q'})$ where ${\hat q}$ is the 
component is the component of the relative momentum $q = (p_1- p_2)/2$ 
{\it orthogonal} to the total hadron 4-momentum 
$P = p_1 + p_2$, then, as was shown in [8], the 4D hadron-quark vertex 
function $\Gamma$ is a function of ${\hat q}$ 
only, and is expressible as ${\Gamma}({\hat q}) = D({\hat q})\phi({\hat q})$, 
where $D$ and $\phi$ are the respective denominator 
and wave functions of the BSE in 3D form, viz., 
$D {\phi} = \int K \phi \ d{\hat q}$. For this 2-body case the 4D and 3D 
forms of the BSE are exactly reversible without further assumptions. For 
the 3-body case a corresponding 3D-4D connection was obtained on the basis 
of semi-intuitive arguments [12], and therefore needed a more formal 
derivation, which is the central aim of this paper. To that end it is useful 
to formulate the 4D and 3D BSE's in terms of Green's functions. Therefore in 
this section we first outline the derivation of the 3D-4D connection for 
a two-body system in terms of their respective Green's functions, in 
preparation for the generalization to the three-body case in the next 
two sections.

	Apart from certain obvious notations which should be clear from 
the context, we shall use the notation and phase conventions of [8,12] for
the various quantities (momenta, propagators, etc). The 4D $qq$ Green's 
function $G(p_1p_2 ; {p_1}'{p_2}')$ near a {\it bound} state satisfies a 
4D BSE without the inhomogeneous term, viz. [8,12],
\setcounter{equation}{0}
\renewcommand{\theequation}{2.\arabic{equation}}
\begin{equation}
i(2\pi)^4 G(p_1 p_2;{p_1}'{p_2}') = {\Delta_1}^{-1} {\Delta_2}^{-1} \int
d{p_1}'' d{p_2}'' K(p_1 p_2;{p_1}''{p_2}'') G({p_1}''{p_2}'';{p_1}'{p_2}')    
\end{equation}
where
\begin{equation}
\Delta_1 = {p_1}^2 + {m_q}^2 ; (m_q = mass of each quark)
\end{equation}
Now using the relative $q = (p_1-p_2)/2$ and total $P = p_1 + p_2$ 
4-momenta (similarly for the other sets), and removing a $\delta$-function
for overall 4-momentum conservation, from each of the $G$- and $K$- 
functions, eq.(2.1) reduces to the simpler form    
\begin{equation}
i(2\pi)^4 G(q.q') = {\Delta_1}^{-1} {\Delta_2}^{-1}  \int d{\hat q}'' 
Md{\sigma}'' K({\hat q},{\hat q'}) G(q,q')
\end{equation}
where ${\hat q}_{\mu} = q_{\mu} - {\sigma} P_{\mu}$, with 
$\sigma = (q.P)/P^2$, is effectively 3D in content (being orthogonal to
$P_{\mu}$). Here we have incorporated the ansatz of a 3D support for the
kernel $K$ (independent of $\sigma$ and ${\sigma}'$), and broken up the 
4D measure $dq''$ arising from (2.1) into the product 
$d{\hat q}''Md{\sigma}''$ of a 3D and a 1D measure respectively. We have 
also suppressed the 4-momentum $P_{\mu}$ label, with $(P^2 = -M^2)$, in 
the notation for $G(q.q')$.
	Now define the fully 3D Green's function ${\hat G}({\hat q},{\hat q}')$
as [8,12] 
\begin{equation}
{\hat G}({\hat q},{\hat q}') = \int \int M^2 d{\sigma}d{\sigma}'G(q,q')
\end{equation}
and two (hybrid) 3D-4D Green's functions ${\tilde G}({\hat q},q')$,
${\tilde G}(q,{\hat q}')$ as
\begin{equation}
{\tilde G}({\hat q},q') = \int Md{\sigma} G(q,q');
{\tilde G}(q,{\hat q}') = \int Md{\sigma}' G(q,q');
\end{equation} 
Next, use (2.5) in (2.3) to give    
\begin{equation}
i(2\pi)^4 {\tilde G}(q,{\hat q}') = {\Delta_1}^{-1} {\Delta_2}^{-1} 
\int dq'' K({\hat q},{\hat q}''){\tilde G}(q'',{\hat q})  
\end{equation}
Now integrate both sides of (2.3) w.r.t. $Md{\sigma}$ and use the result [8]
\begin{equation}
\int Md{\sigma}{\Delta_1}^{-1} {\Delta_2}^{-1} = 2{\pi}i D^{-1}({\hat q});
D({\hat q}) = 4{\hat \omega}({\hat \omega}^2 - M^2/4);
{\hat \omega}^2 = {m_q}^2 + {\hat q}^2 
\end{equation}
to give a 3D BSE w.r.t. the variable ${\hat q}$, while keeping the other 
variable $q'$ in a 4D form:
\begin{equation}  
(2\pi)^3 {\tilde G}({\hat q},q') = D^{-1} \int d{\hat q}''  
K({\hat q},{\hat q}'') {\tilde G}({\hat q}'',q')
\end{equation}
Now a comparison of (2.6) with (2.8) gives the desired connection between 
the full 4D $G$-function: 
\begin{equation}  
2{\pi}i G(q,q') = D({\hat q}){\Delta_1}^{-1}{\Delta_2}^{-1}
{\tilde G}({\hat q},q')
\end{equation}
which is the Green's function counterpart, {\it{near the bound state}}, 
of the same result [8] connecting the corresponding BS wave functions.
Again, the symmetry of the left hand side of (2.9) w.r.t. $q$ and $q'$ 
allows us to write the right hand side with the roles of $q$ and $q'$ 
interchanged. This gives the dual form   
\begin{equation}  
2{\pi}i G(q,q') = D({\hat q}'){{\Delta_1}'}^{-1}{{\Delta_2}'}^{-1}
{\tilde G}(q,{\hat q}')
\end{equation}
which on integrating both sides w.r.t. $M d{\sigma}$ gives
\begin{equation}  
2{\pi}i{\tilde G}({\hat q},q') = D({\hat q}'){{\Delta_1}'}^{-1}
{{\Delta_2}'}^{-1}{\hat G}({\hat q},{\hat q}'). 
\end{equation}
Substitution of (2.11) in (2.9) then gives the symmetrical form
\begin{equation}  
(2{\pi}i)^2 G(q,q') = D({\hat q}){\Delta_1}^{-1}{\Delta_2}^{-1}
{\hat G}({\hat q},{\hat q}')D({\hat q}'){{\Delta_1}'}^{-1}
{{\Delta_2}'}^{-1}
\end{equation}
Finally, integrating both sides of (2.8) w.r.t. $M d{\sigma}'$, we 
obtain a fully reduced 3D BSE for the 3D Green's function:
\begin{equation}  
(2\pi)^3 {\hat G}({\hat q},{\hat q}') = D^{-1}({\hat q} \int d{\hat q}''
K({\hat q},{\hat q}'') {\hat G}({\hat q}'',{\hat q}')
\end{equation}
Eq.(2.12) which is valid near the bound state pole (since the 
inhomogeneous term has been dropped for simplicity) expresses the desired 
connection between the 3D and 4D forms of the Green's functions; and 
eq(2.13) is the determining equation for the 3D form. A spectral analysis 
can now be made for either of the 3D or 4D Green's functions in the 
standard manner, viz., 
\begin{equation}  
G(q,q') = \sum_n {\Phi}_n(q;P){\Phi}_n^*(q';P)/(P^2 + M^2) 
\end{equation}
where $\Phi$ is the 4D BS wave function. A similar expansion holds for 
the 3D $G$-function ${\hat G}$ in terms of ${\phi}_n({\hat q})$. Substituting
these expansions in (2.12), one immediately sees the connection between 
the 3D and 4d wave functions in the form:
\begin{equation}  
2{\pi}i{\Phi}(q,P) = {\Delta_1}^{-1}{\Delta_2}^{-1}D(\hat q){\phi}(\hat q)
\end{equation}
whence the BS vertex function becomes $\Gamma = D \times \phi/(2{\pi}i)$
as found in [8]. We shall make free use of these results, taken as $qq$ 
subsystems, for our study of the $qqq$ $G$-functions in Sections 3 and 4.  

\section{Three-Quark Green's Function: 3D Reduction of the BSE}

	As in the two-body case, and in an obvious notation for various 
4-momenta (without the Greek suffixes), we consider the most general 
Green's function $G(p_1 p_2 p_3;{p_1}' {p_2}' {p_3}')$ for 3-quark 
scattering {\it near the bound state pole} (for simplicity) which allows           
us to drop the various inhomogeneous terms from the beginning. Again we 
take out an overall delta function $\delta(p_1 + p_2 + p_3 - P)$ from the
$G$-function  and work with {\it two} internal 4-momenta for each of the 
initial and final states defined as follows [12]:

\setcounter{equation}{0}
\renewcommand{\theequation}{3.\arabic{equation}}

\begin{equation}  
{\sqrt 3}{\xi}_3 =p_1 - p_2 \ ; \quad  3{\eta}_3 = - 2p_3 + p_1 +p_2
\end{equation}
\begin{equation}  
P = p_1 + p_2 + p_3 = {p_1}' + {p_2}' + {p_3}'
\end{equation}
and two other sets ${\xi}_1,{\eta}_1$ and ${\xi}_2,{\eta}_2$ defined by 
cyclic permutations from (3.1). Further, as we shall be considering pairwise
kernels with 3D support, we define the effectively 3D momenta ${\hat p}_i$, 
as well as the three (cyclic) sets of internal momenta 
${\hat \xi}_i,{\hat \eta}_i$, (i = 1,2,3) by [12]:
\begin{equation}  
{\hat p}_i = p_i - {\nu}_i P \ ;\quad  {\hat {\xi}}_i = {\xi}_i - s_i P\  ;
\quad
{\hat {\eta}}_i - t_i P 
\end{equation}
\begin{equation}  
\\nu_i = (P.p_i)/P^2\  ;\quad s_i = (P.\xi_i)/P^2 \ ;\quad t_i = 
(P.\eta_i)/P^2 \end{equation}
\begin{equation}  
{\sqrt 3} s_3 = \nu_1 - \nu_2 \ ;\quad 3 t_3 = -2 \nu_3 + \nu_1 + \nu_2 \ 
\ ( + {\rm cyclic permutations})
\end{equation}

The space-like momenta ${\hat p}_i$ and the time-like ones $\nu_i$ 
satisfy [12] 
\begin{equation}  
{\hat p}_1 + {\hat p}_2 + {\hat p}_3 = 0\  ;\quad \nu_1 + \nu_2 + \nu_3 = 1
\end{equation}
Strictly speaking, in the spirit of covariant instantaneity, we should 
have taken the relative 3D momenta ${\hat \xi},{\hat \eta}$ to be in the 
instantaneous frames of the concerned pairs, i.e., w.r.t. the rest frames
of $P_{ij} = p_i +p_j$; however the difference between the rest frames of 
$P$ and $P_{ij}$  is small and calculable [12], while the use of a common 
3-body rest frame $(P = 0)$ lends considerable simplicity and elegance to 
the formalism.   
	We may now use the foregoing considerations to write down the BSE 
for the 6-point Green's function i terms of relative momenta, on closely 
parallel lines to the 2-body case. To that end note that the 2-body 
relative momenta $q_{ij} = (p_i - p_j)/2 = {sqrt 3}{\xi_k}/2$, where 
(ijk) are cyclic permutations of (123). Then for the reduced $qqq$ Green's
function, when the {\it last} interactio was in the (ij) pair, we may use 
the notation $G(\xi_k \eta_k ; {\xi_k}' {\eta_k}')$, together with 'hat' 
notations on these 4-momenta when the corresponding time-like components 
are integrated out. Further, since the pair $\xi_k,\eta_k$ is 
{\it {permutation invariant}} as a whole, we may choose to drop the index 
notation from the complete $G$-function to emphasize this symmetry as and 
when needed. The $G$-function for the $qqq$ system satisfies, in the 
neighbourhood of the bound state pole, the following (homogeneous) 4D BSE
for pairwise $qq$ kernels with 3D support:
\begin{equation}  
i(2\pi)^4 G(\xi \eta ;{\xi}' {\eta}') = \sum_{123}
{\Delta_1}^{-1} {\Delta_2}^{-1} \int d{{\hat q}_{12}}'' M d{\sigma_{12}}''
K({\hat q}_{12}, {{\hat q}_{12}}'') G({\xi_3}'' {\eta_3}'';{\xi_3}' {\eta_3}')  
\end{equation}  
where we have employed a mixed notation ($q_{12}$ versus $\xi_3$) to stress
the two-body nature of the interaction with one spectator at a time, in a 
normalization directly comparable with eq.(2.3) for the corresponding 
two-body problem. Note also the connections 
\begin{equation}  
\sigma_{12} = {\sqrt 3}{s_3}/2 \  ;\quad 
{\hat q}_{12} = {\sqrt 3}{{\hat \xi}_3}/2 \ ; \quad \eta_3 = - p_3, etc 
\end{equation}  
The next task is to reduce the 4D BSE (3.7) to a fully 3D form through a 
sequence of integrations w.r.t. the time-like momenta $s_i,t_i$ applied 
to the different terms on the right hand side, {\it {provived both}} 
variables are simultaneously permuted. We now define the following fully 
3D as well as mixed 3D-4D $G$-functions according as one or more of the 
time-like $\xi,\eta$ variables are integrated out:
\begin{equation}  
{\hat G}({\hat \xi} {\hat \eta};{\hat \xi}' {\hat \eta}') = 
\int \int \int \int ds dt ds' dt' G(\xi \eta ; {\xi}' {\eta}')  
\end{equation}  
which is $S_3$-symmetric.
\begin{equation}  
{\tilde G}_{3\eta}(\xi {\hat \eta};{\xi}' {\hat \eta}') = 
\int \int dt_3 d{t_3}' G(\xi \eta ; {\xi}' {\eta}');
\end{equation}  
\begin{equation}  
{\tilde G}_{3\xi}({\hat \xi}  \eta;{\hat \xi}' {\eta}') = 
\int \int ds_3 d{s_3}' G(\xi \eta ; {\xi}' {\eta}');
\end{equation} 
The last two equations are however $S-3$-indexed. The full 3D BSE for the 
${\hat G}$-function is obtained by integrating out both sides of (3.7)
w.r.t. $ds_i d{s_j}' dt_i d{t_j}'$ ($S_3$-symmetric), and using (3.9) with
(3.8) as follows:
\begin{equation}  
(2\pi)^3 {\hat G}({\hat \xi} {\hat \eta} ;{\hat \xi}' {\hat \eta}') = 
\sum_{123} D^{-1}({\hat q}_{12}) \int d{{\hat q}_{12}}'' 
K({\hat q}_{12}, {{\hat q}_{12}}'') {\hat G}({\hat \xi}'' {\hat \eta}'';
{\hat \xi}' {\hat \eta}')  
\end{equation}   
This integral equation for ${\hat G}$ which is the 3-body counterpart of
(2.13) for a $qq$ system in thev neighbourhood of the bound state pole, 
is the desired 3D BSE for the $qqq$ system in a {\it {fully connected}}
form, i.e., free from delta functions. Now using a spectral decomposition 
for ${\hat G}$ 
\begin{equation}   
{\hat G}({\hat \xi} {\hat \eta};{\hat \xi}' {\hat \eta}')
= \sum_n {\phi}_n( {\hat \xi} {\hat \eta} ;P)
{\phi}_n^*({\hat \xi}' {\hat \eta}';P)/(P^2 + M^2)
\end{equation}   
on both sides of (3.12) and equating the residues near a given pole
$P^2 = -M^2$, gives the desired equation for the 3D wave function $\phi$ 
for the bound state in the connected form:
\begin{equation}   
(2\pi)^3 \phi({\hat \xi} {\hat \eta} ;P) = \sum_{123} D^{-1}({\hat q}_{12})
\int d{{\hat q}_{12}}'' K({\hat q}_{12}, {{\hat q}_{12}}'')
\phi({\hat \xi}'' {\hat \eta}'' ;P)
\end{equation}   
The solution of this equation for the ground state was found in [11] in 
a {\it gaussian} form which implies that $\phi({\hat \xi} {\hat \eta};P)$ 
is an $S_3$-invariant function of ${{\hat \xi}_i}^2 + {{\hat \eta}_i}^2$, 
{\it {valid for any index i}}. While the gaussian form may prove too 
restrictive for more general applications, the mere $S_3$-symmetry of 
$\phi$ in the $({\hat \xi}_i, {\hat \eta}_i)$ pair may prove adequate 
in practice, and hence useful for both the solution of (3.14) {\it and}
for the reconstruction of the 4D BS wave function in terms of the 3D 
wave function (3.14), as is done in Sec.4 below.

\section{Reconstruction of the 4D BS Wave Function}
	In this section we shall attempt to {\it re-express} the 4D 
$G$-function given by (3.7) in terms of the 3D ${\hat G}$-function 
given by (3.12), as the $qqq$ counterpart of the $qq$ results (2.12-13). 
To that end we first adapt the result (2.12) to the hybrid Green's 
function  of the (12) subsystem given by ${\tilde G}_{3 \eta}$, 
eq.(3.10), in which the 3-momenta $\eta_3,{\eta_3}'$ play a parametric 
role reflecting the spectator status of quark $\# 3$, while the {\it active} 
roles are played by $q_{12},{q_{12}}' = {\sqrt 3}(\xi_3, {\xi_3}')/2$, 
for which the analysis of Sec.2 applies directly. This gives    

\setcounter{equation}{0}
\renewcommand{\theequation}{4.\arabic{equation}}
\begin{equation}
(2{\pi}i)^2 {\tilde G}_{3 \eta}(\xi_3 {\hat \eta}_3; 
{\xi_3}' {{\hat \eta}_3}') 
= D({\hat q}_{12}){\Delta_1}^{-1}{\Delta_2}^{-1}
{\hat G}({\hat \xi_3} {\hat \eta_3}; {\hat \xi_3}' {\hat \eta_3}')
D({{\hat q}_{12}}'){{\Delta_1}'}^{-1}{{\Delta_2}'}^{-1}
\end{equation}
where on the right hand side, the `hatted' $G$-function has full 
$S_3$-symmetry, although (for purposes of book-keeping)we have not 
shown this fact explicitly by deleting the suffix `3' from its 
arguments. A second relation of this kind mau be obtained from (3.7)
by noting that the 3 terms on its right hand side may be expressed in 
terms of ${\tilde G}_{3 \xi}$ functions vide their definitions (3.11), 
together with the 2-body interconnection between $(\xi_3,{\xi_3}')$ 
and $({\hat \xi}_3,{{\hat \xi}_3}')$ expressed once again via (4.1), 
but without the `hats' on $\eta_3$ and ${\eta_3}'$. This gives
\begin{eqnarray}
({\sqrt 3} \pi i)^2 G(\xi_3 \eta_3; {\xi_3}'{\eta_3}')
&=& ({\sqrt 3} \pi i)^2 G(\xi \eta; {\xi}'{\eta}')\nonumber\\
&=& \sum_{123} {\Delta_1}^{-1}{\Delta_2}^{-1} (\pi i {\sqrt 3})
\int d{{\hat q}_{12}}'' M d{\sigma_{12}}''
K({\hat q}_{12}, {{\hat q}_{12}}'') 
G({\xi_3}'' {\eta_3}'';{\xi_3}' {\eta_3}')\nonumber\\   
&=& \sum_{123} D({\hat q}_{12}) {\Delta_1}^{-1}{\Delta_2}^{-1}
{\tilde G}_{3 \xi}({\hat \xi}_3  \eta_3; {{\hat \xi}_3}' {{\eta}_3}')
{{\Delta_1}'}^{-1} {{\Delta_2}'}^{-1}  
\end{eqnarray}
where the second form exploits the symmetry between $\xi,\eta$ and 
$\xi',\eta'$.
	This is as far as we can go with the $qqq$ Green's function, 
using the 2-body techniques of Sec.2. However,, unlike the 2-body case 
where the reconstruction of the 4D $G$-function in terms of the 
corresponding 3D quantity was complete at this stage, the process is
far from complete for the 3-body case, as eq.(4.2) clearly shows. This
is due to the {\it truncation} of Hilbert space implied in the ansatz 
of 3d support to the pairwise BSE kernel $K$ which, while facilitating a 
4d to 3d BSE reduction without extra charge, does {\it not} have the 
{\it complete} information to permit the {\it reverse} transition 
(3d to 4D) without additional assumptions. This limitation of the 3D 
support ansatz for the BSE kernel affects all $n$-body systems except
$n = 2$ (which may be regarded as a sort of degenerate situation.
	Now it may be argued: Is this 3D ansatz for the BSE kernel 
really necessary? Since this paper is not the most convenient place 
to dwell on this {\it physical} issue in detail (the same has been 
discussed elsewhere [10], vis-a-vis contemporary approaches), we 
shall here take the 3D support ansatz for granted, and look upon 
this "inverse" problem as a purely {\it mathematical} one. We add in 
parentheses however that the physical applications of the 3D ansatz 
are (indeed) quite widespread since it is directly related to the 
"instantaneous approximation" on which a considerable amount of low 
and intermediate energy hadron physics (at the quark level) has been 
(and is still being) studied.
	As a purely mathematical problem, we must look for a suitable
ansatz for the quantity ${\tilde G}_{3 \xi}$ on the right hand side of 
(4.2) in terms of {\it known} quantities, so that the reconstructed 
4D $G$-function satisfies the 3D equation (3.12) exactly, which may 
be regarded as a "check-point" for the entire exercise. We therefore
seek a structure of the form 
\begin{equation}
{\tilde G}_{3 \xi}({\hat \xi}_3  {\eta}_3; {{\hat \xi}_3}' {{\eta}_3}')
= {\hat G}({{\hat \xi}_3} {\hat \eta}_3; {{\hat \xi}_3}' {{\hat \eta}_3}')
\times F(p_3, {p_3}')    
\end{equation}
where the unknown function $F$ must involve only the momentum of the 
spectator quark $\# 3$. A part of the $\eta_3, {\eta_3}'$ dependence has 
been absorbed in the ${\hat G}$ function on the right, so as to satisfy 
the requirements of $S_3$-symmetry for this 3D quantity, whether it has 
a gaussian structure [11] (where it is explicit), or a more general 
one; see the discussion below eq(3.14).
	As to the remaining factor $F$, it is necessary to choose its
form in a careful manner so as to conform to the conservation of 
4-momentum for the {\it free} propagation of the spectator between two
neighbouring vertices, consistently with the symmetry between $p_3$ 
and ${p_3}'$. A possible choice consistent with these conditions is
the form
\begin{equation}
F(p_3, {p_3}') = C_3 {\Delta_3}^{-1} {\delta}(\nu_3 - {\nu_3}') 
\end{equation}
where $\Delta_3$ could also be written more symmetrically as 
${\sqrt \Delta_3 {\Delta_3}'}$.
	Here we have taken only the time component of the 4-momentum 
$p_3$ in the $\delta$-function since the effect of its space component 
has already been absorbed in the "connected" (3D) Green's function 
${\hat G}$. ${\Delta_3}^{-1}$ represents the "free" propagation of 
quark $\# 3$ between successive vertices, while $C_3$ represents some 
residual effects which may at most depend on the 3-momentum 
${\hat p}_3$, but must satisfy the main constraint that the 3D BSE, 
eq.(3.12), is ${\it {explicitly satisfied}}$.
	To check the self-consistency of the ansatz (4.4), integrate
both sides of (4.2) w.r.t. $ds_3 d{s_3}' dt_3 d{t_3}'$ to recover the 
3D $S_3$-invariant ${\hat G}$-function on the left hand side; and in 
the first form on the right hand side, integrate w.r.t. $ds_3 d{s_3}'$ 
on the $G$-function which alone involves these variables. This yields
the quantity ${\tilde G}_{3 \xi}$. At this stage, employ the ansatz 
(4.4) to integrate over $dt_3 d{t_3}'$. Consistency with the 3D BSE, 
eq.(3.12), now demands 
\begin{equation}
C_3 \int \int d\nu_3 d{\nu_3}' {\Delta_3}^{-1} \delta(\nu_3 - {\nu_3}')
= 1 ; (since dt = d\nu) 
\end{equation}
The 1D integration w.r.t. $d\nu_3$ may be evaluated as a contour 
integral over the propagator ${\Delta}^{-1}$ , which gives the pole 
at $\nu_3 = {\hat \omega}_3/M$, (see below). Evaluating the residue
then gives 
\begin{equation}
C_3 = i \pi / (M {\hat \omega}_3 ) ;  \quad
{{\hat \omega}_3}^2 = {m_q}^2 + {{\hat p}_3}^2
\end{equation}
which will reproduce the 3D BSE, eq.(3.12), {\it exactly}! Substitution
of (4.4) in the second form of (4.2) finally gives the desired 3-body 
generalization of (2.12) in the form 
\begin{equation}
3 G(\xi \eta; \xi' \eta') = \sum_{123} D({\hat q}_{12}) \Delta_{1F} 
\Delta_{2F} D({{\hat q}_{12}}') {\Delta_{1F}}' {\Delta_{2F}}' 
{\hat G}({\hat \xi_3} {\hat \eta_3}; {\hat \xi_3}' {\hat \eta_3}')
[\Delta_{3F} / (M \pi {\hat \omega}_3)]      
\end{equation}
where for each index, $\Delta_F = - i {\Delta}^{-1}$ is the 
Feynman propagator.
	Before commenting on this structure of the 4D Green's function 
near the bound state pole, let us first find the effect of the ansatz 
(4.4) on the 4D BS {\it {wave function}} $\Phi(\xi \eta; P)$. This is 
achieved through a spectral representation like (3.13) for the 4D 
Green's function $G$ on the left hand side of (4.2). Equating the 
residues on both sides gives the desired 4D-3D connection between 
$\Phi$ and $\phi$:
\begin{equation}
\Phi(\xi \eta; P) = \sum_{123} D({\hat q}_{12}){\Delta_1}^{-1}{\Delta_2}^{-1}
\phi ({\hat \xi} {\hat \eta}; P) \times 
\sqrt{{\delta(\nu_3 -{\hat \omega}_3/M)} \over{M {\hat \omega}_3 
{\Delta}_3}} 
\end{equation}
From (4.8) we can infer the structure of the baryon-$qqq$ vertex function 
by rewriting the it in the alternative form [12]:
\begin{equation}
\Phi(\xi \eta; P) = (V_1 + V_2 + V_3) \over {\Delta_1 \Delta_2 \Delta_3}
\end{equation}
where 
\begin{equation}
V_3 = D({\hat q}_{12})\phi ({\hat \xi} {\hat \eta}; P) \times
\sqrt{{\Delta}_3 \delta (\nu_3 -{\hat \omega}_3/M) \over{M {\hat \omega}_3}}
\end{equation}
The quantity $V_3$ is the baryon-$qqq$ vertex function corresponding to 
the "last interaction" in the (12) pair, and so on cyclically. This is 
precisely the form (apart from a constant factor that does not affect
the baryon normalization) that had been anticipated in an earlier
study in a semi-intuitive fashion [12].

\section{Discussion and Conclusion}
 
	Eqs.(4.7-10) which represent the principal results of this 
investigation, bring out rather directly the significance of the 
square root of the $\delta$-function in the energy variable of the 
spectator. Both the $\delta$-funnction and the $\Delta_{3F}$ propagator 
appear in {\it {rational forms in the 4D Green's function}}. reflecting 
a free on-shell propagation of the spectator between two vertex points.
The square root feature in the baryon-$qqq$ vertex function is the 
result of equal distribution of this singulariity between the initial
and final state vertex functions. Further, as the steps indicate, the
appearace of this singularity has nothing to do with the connectedness
[3,4] of the 3-body scattering amplitude, but rather with the 3D support
for the pairwise BSE kernel. More importantly, this singularity will 
{\it not} show up in any physical amplitude for hadronic transitions
via quark loops, since such amplitudes will always involve both the 
$\delta$-function and the propagator $\Delta_{3F}$ in a {\it rational} 
form before the relevant momentum integrationns are performed.
	The next question concerns the possible uniqueness of the 
structure (4.7-10). It is certainly "sufficient" since the 3D form 
(3.12) of the BSE for the ${\hat G}$- function is exactly satisfied.
Moreover the underlying ansatz (4.4) has certain desirable properties
like on-shell propagation of the spectator in between two successive 
interactions, as well as an explicit symmetry in the $p_3$ and ${p_3}'$
momenta. There is a fair chance of its uniqueness within certain 
general constraints, but so far we have not been able to prove this.
	As regards spin, the extension of the above formalism to 
fermion quarks is a straightforward process amounting to the 
replacement of $\Delta_F = -i \Delta^{-1}$ by the corresponding 
$S_F$-functions, as has been shown elsewhere [11,12]. In particular, 
the fermion vertex function has recently been applied to the problem
of proton-neutron mass difference [13] via quark loop integrals, to 
bring out the practicability of its application without parametric 
uncertainties, since the entire formalism is linked all the way from 
spectroscopy to hadronic transition amplitudes [7].
	At this stage it is interesting to ask, again as a purely 
mathematical problem, what would have been the possible scenario 
if the 3D support ansatz for the BSE kernel had not been made.
In that case, the entire "hat" formalism would become redundant,
and there would be no special roles of equations like (2.12) or 
(4.7) connecting the 4D to the 3D Green's functions. The "connected"  
equations (2.13) or (3.12) would simply remain valid {\it without} 
the "hats", viz., as 4D integral equations, and with the replacement 
of $D({\hat q}_{12})$ by $\Delta_1 \Delta_2/(2i \pi)$. And if closed 
form solutions of these equations were routinely possible in 4D form,
there would  be no special advantage in going in for more 
complicated connected equations [3,4]. This kind of 3-body approach in 
direct 4D form was indeed attempted in a Wick-rotated Euclidean manner
[5], but the predicted O(4)-like spectra did not accord with observation 
[6]. And while `normal' 4D kernels (i.e., without Wick rotation) have 
been employed for ${\bar q}q$ systems [14], there is no corresponding 
evidence of the $qqq$ BSE attempted on similar lines. The 3D kernel 
support discussed here is just a concrete alternative  which, being 
otherwise rooted in a 4D framework, acts as an effective bridge between 
traditional 3D methods employed in the literature  for few-body problems 
at the quark level and more formal 4D treatments [14-17]; ( see next para 
for specific comparison with other $qqq$ problem studies). Its strong 
connection with the `instantaneous approximation'  gives it a natural 
applicational base for a  systematic treatment of hadronic phenomena at 
the quark level up to moderately high energies, as indicated by its 
observational successes [8,10-13]. 
	Finally we would like to comment on this formalism vis-a-vis 
contemporary approaches to the $qqq$ problem. Many such approaches as are 
available in the literature are parametric representations attuned to 
QCD-sum rules [15], effective Lagrangians for hadronic transitions to
"constituent" quarks, with ad hoc assumptions on the hadron-$qqq$ form 
factor [16], similar (parametric) ansatze for the hadron- quark-diquark 
form factor [17]; or more often simply direct gaussian parametrizations 
for the $qqq$ wave functions as the starting point of the investigation 
[18]. Such approaches are often quite effective for the investigations of
some well-defined sectors of hadron physics with quark degrees of freedom,
but are not readily extensible to other sectors without further 
assumptions ( e.g., meson and baryon parametrizations are quite unrelated 
to each other), since there is no possibility of a deeper understanding of
the crucial functions/parameters involved, from more formal dynamical 
premises. A more unified approach, albeit at the cost of a bigger dynamical
investment, should have the capacity to provide a more natural form of 
integration of the different sectors, perhaps all the way to hadron 
spectroscopy, without additional assumptions on the way. Such 
approaches  usually need a "dynamical equation" such as the BSE or SDE, 
as the starting point for the flow of information. It is precisely in 
respect of such unification that the philosophy underlying the present 
formalism for the $qqq$ problem differs from some others [15-18]. This 
is hardly a new philosophy, since the perspective in this respect was 
shown 25 years ago by Feynman [19], but can stand a reiteration.
	This work arose out of the need for a formal demonstration of a 
semi-intuitive ansatz [12] on the structure of the baryon-$qqq$ vertex 
function that had been recently applied to the neutron-proton mass 
difference problem [13], on the demand of some critics. However it 
took shape as a self-contained "mathematical" problem in its own 
right, even though its origin is strongly physical. 
	The final version of this paper was prepared at the International 
Centre for Theoretical Physics  during a short time visit of the author
in November 1996. He expresses his appreciation to the Director, Prof.
M.A.Virasoro, for this hospitality.

\end{document}